\documentclass[letterpaper,twocolumn,english,aps,pre]{revtex4}
\usepackage{times}
\usepackage[T1]{fontenc}
\usepackage[latin1]{inputenc}
\usepackage{color}
\usepackage{graphicx}

\makeatletter

\usepackage{hyperref}
\newcommand{\NPR}{N_{\mathrm{PR}}}

\usepackage{babel}
\makeatother
\begin{document}

\title{Racing downhill: optimization and the random-field Ising model}

\author{D. Clay Hambrick}

\affiliation{Department of Physics, Harvey Mudd College, Claremont, CA 91711}

\affiliation{Department of Physics, Syracuse University, Syracuse, NY 13244}

\author{Jan H. Meinke}

\author{A. Alan Middleton}

\affiliation{Department of Physics, Syracuse University, Syracuse, NY 13244}

\date{January 10, 2005}

\begin{abstract}
The push-relabel algorithm can be used to calculate rapidly the exact
ground states for a given sample with a random-field Ising model (RFIM)
Hamiltonian. Although the algorithm is guaranteed to terminate after
a time polynomial in the number of spins, implementation details are
important for practical performance. Empirical results for the timing
in dimensions $d=1,2,\textrm{ and }3$ are used to determine the fastest
among several implementations. Direct visualization of the auxiliary
fields used by the algorithm provides insight into its operation and
suggests how to optimize the algorithm. Recommendations are given
for further study of the RFIM. 
\end{abstract}
\maketitle

\section{Introduction}

In systems with quenched disorder, changes in the random background
take place over time scales that are much longer than the time scale
for evolution of the primary degrees of freedom. In magnetic systems,
the spin degrees of freedom interact in an effectively frozen random
environment determined by substitutional disorder or vacancies. The
random-field Ising model (RFIM), defined as ferromagnetically-coupled
spins subject to a spatially varying magnetic field, is a prototypical
model for magnets with quenched disorder that has been studied since
the 1970s (for reviews, see, e.g., \cite{YoungBook}). In dimensions
$d>2$, the RFIM has a transition between ferromagnetic (FM) and paramagnetic
(PM) states; this transition can be found by varying either the temperature
or the disorder, at sufficiently small values of the remaining control
parameter. Fishman and Aharony \cite{Fishman1979} mapped the RFIM
with a field of random sign and fixed magnitude with disordered bonds
to an experimentally realizable system: the diluted antiferromagnet
in a field (DAFF) \cite{Belanger1998}. At low temperatures, the glassy
behavior of the DAFF seen in both experiment and theory leads to non-equilibrium
effects such as history dependence and a broad range of relaxation
times. These observations are qualitatively consistent with predictions
by both Fisher and Villain of an exponential slowing down near the
critical point and of a low temperature phase described by the zero-temperature
critical point \cite{Villain1984,Fisher1986}.

Exponential slowing down also affects optimization methods such as
simulated annealing \cite{Kirkpatrick83}, that are modeled on the
dynamics of the physical system. The large barriers to equilibration
make it very time consuming to sample configuration space accurately
at finite temperatures or to find the exact zero-temperature ground
states using such methods. Finding the partition function for the
RFIM at finite temperature is NP-hard \cite{JCAd85,SI99}. However,
the zero-temperature FM-PM transition is expected to be in the same
universality class as the finite temperature transition. So it is
fortunate that there are alternate methods for quickly finding the
exact RFIM ground state. These methods are based on a mapping from
the problem of finding the ground state of the RFIM to that of finding
the maximum flow---or, equivalently, the minimum cut---in a capacitated
graph \cite{Picard1975,Barahona1985,Ogielski1986}. Ogielski's computations
of the RFIM ground state properties \cite{Ogielski1986} utilized
the push-relabel (PR) algorithm for max-flow introduced by Goldberg
and Tarjan \cite{Goldberg1988}. The PR algorithm is in practice the
most efficient algorithm available for many classes of problems \cite{Cherkassky1997}.
Although its generic implementation has a polynomial time bound, its
actual performance depends on the order in which operations are performed
and which heuristics are used to maintain auxiliary fields for the
algorithm. Even within this polynomial time bound, there is a power-law
critical slowing down of the PR algorithm at the zero-temperature
($T=0$) transition \cite{Ogielski1986,MiddletonFisherRFIM,MiddletonCritSlow2002}.

This paper presents results that are useful for minimizing CPU time
in RFIM ground-state simulations. We begin with a brief review of
the RFIM, including its definition and phases, in Sec.~\ref{sec:Random-field-Ising}.
We then discuss the implementations for the PR algorithms that we
study. The PR algorithm redistributes the random magnetic field among
spins by pushing positive field {}``downhill'' with respect to a
potential field (the {}``height'') defined for each spin. This redistribution
and coalescence of positive field and negative field {}``sinks''
allows for the determination of same-spin domains. The data structures
used to organize the pushes and the heuristics used for updating the
potential field are described in Sec.~\ref{sec:Generic-Push-Relabel-Algorithm.}.
The results for the timing using various heuristics, summarized in
Sec.~\ref{sec:Timing-Results}, should be useful for designing further
extensive studies of the RFIM. Visualizing the auxiliary fields leads
to a clearer explanation of the timing results and was important in
guiding our work. In Sec.~\ref{sec:Qualitative-Description}, we
use these visualizations to present a qualitative overview of the
operations of the different implementations in the distinct phases
of the RFIM. The primary results of the paper, namely the recommended
choices for the PR algorithm, are summarized in Sec.~\ref{sec:Summary}.

\section{Random-field Ising model\label{sec:Random-field-Ising}}

The random-field Ising model has a non-trivial ground state due to
the competition between the ferromagnetic interaction that tends to
align neighboring spins and the influence of random fields, which
tends to force spins to point in random directions. Taking the strength
of the ferromagnetic interactions between neighboring spins to be
$J$ and designating the local random fields by $h_{i}$, the energy
of a spin configuration is \cite{YoungBook}\begin{equation}
\mathcal{H}=-J\sum_{\langle ij\rangle}s_{i}s_{j}-\sum_{i}h_{i}s_{i},\label{eq:RFIM_Hamiltonian}\end{equation}
 where the spin $s_{i}$ at a given site $i$ on a $d$-dimensional
lattice with $n=L^{d}$ sites can take on values $s_{i}=\pm1$ and
the sites $i$ and $j$ in the sum in the first term are nearest neighbors.
We study the Gaussian RFIM, where the $h_{i}$ are independent variables
chosen from a Gaussian distribution with mean $0$ and variance $\Delta^{2}J^{2}$,
with periodic boundary conditions. The parameter $\Delta$ characterizes
the strength of the disorder relative to the ferromagnetic interaction.

As the disorder dominates over thermal fluctuations at large length
scales and low temperature in more than two dimensions, the ground
states of this model are of interest. In dimensions $d>2$, there
is a zero-temperature transition between two phases at the critical
disorder $\Delta=\Delta_{c}$. When $\Delta<\Delta_{c}$, the ferromagnetic
interaction between nearest neighbors dominates and the spins take
on a mean value $m=n^{-1}\sum_{i}s_{i}$ with $\left|m\right|\neq0$
in the limit $n\rightarrow\infty$. In the case $\Delta>\Delta_{c}$,
randomness dominates and the ground state is {}``paramagnetic''
$\left|m\right|=0$, as $n\rightarrow\infty$.

In dimensions $d=1\textrm{ and }2$, the ground state is in the paramagnetic
phase at any $\Delta>0$ in the thermodynamic limit (i.e., $\Delta_{c}=0$).
But the correlation length $\xi$, characterizing the range of spin-spin
correlations or the size of uniform spin domains, diverges as $\Delta^{-2}$
when $d=1$ (see, e.g., \cite{SchroderKnetterAlavaRieger2002}). For
samples of size $L\ll\xi$, i.e., for $\Delta<\Delta_{x}\sim L^{-1/2}$,
where $\Delta_{x}(L)$ is the sample-size-dependent crossover disorder,
the ground state is essentially ferromagnetic. It has been argued
that $\ln(\xi)$ scales as an inverse power of $\Delta$ when $d=2$
\cite{Binder1983,GrinsteinMa82,SeppalaAlava2D2001}. This very rapidly
growing correlation length gives rise to an apparent ferromagnetic
phase in $d=2$ at even moderate $\Delta_{x}$ in finite samples \cite{AlavaRieger1998}.
When we take the limit $\Delta\rightarrow0$ in $d=1$ and $d=2$,
we will be taking $\Delta\ll\Delta_{x}$. Many of the results for
the algorithm, such as which algorithm is fastest, hold independently
of the existence of a true physical phase transition.

\section{\label{sec:Generic-Push-Relabel-Algorithm.}Push-Relabel Algorithm:
Data Structures and Heuristics}

We now discuss the motivation for using the PR algorithm and outline
its structure. In particular, we define the auxiliary fields and basic
operations that operate on those fields to determine the ground state.
Picard and Ratliff showed that any quadratic optimization problem,
such as the RFIM, can be mapped onto a min-cut/max-flow problem \cite{Picard1975}.
There are a number of algorithms for solving max-flow \cite{Cormen1990},
though Cherkassky and Goldberg's results show that the PR algorithm
\cite{Goldberg1988} is often the best algorithm for solving large
problems on a variety of graphs \cite{Cherkassky1997}. For detailed
explanations of the mapping of the RFIM to a max-flow problem and
the proofs of the correctness of the PR algorithm, see reviews of
applications of combinatorial optimization to statistical physics
\cite{Alava2001,HartmannRiegerBook2002,Middleton2004} and computer
science texts \cite{Cormen1990}. In this paper, we limit ourselves
to a description of the algorithm, neglecting proofs of correctness,
but including a {}``physical'' description for the variant of the
PR algorithm we have used.

Intuitively, the PR algorithm finds the domains of uniform spin in
the ground state by rearranging ({}``pushing'') the magnetic field.
If the bond between two spins is strong enough, the field on one spin
can be removed from one spin and added to its neighbor, possibly influencing
the direction of the neighboring spin. This rearrangement (and change
in the bonds, as described below) is the push. Subsequent pushes can
then affect distant spins. As a consequence of pushes, positive and
negative fields originally located on separate spins cancel. The cancellation
leads to domains of uniform sign for the excess fields. This domain
growth by rearrangement of field is limited by the strength of the
nearest-neighbor bonds, which {}``carry'' the rearrangement. The
push operation reduces or removes the interactions between neighboring
spins. When the field has large variations compared to the bond strengths,
the magnetic field cannot be pushed very far as bonds become saturated
and block further rearrangement. Conversely, in the limit of weak
fields, large domains form, as the bonds favor alignment of nearest
neighbor spins: in the language of PR, the large capacity of the bonds
relative to the strength of the fields allows for long-range rearrangements
of the random field. In the limit of very weak fields, the field can
be pushed anywhere (no bonds are saturated), so there is only one
domain, whose orientation is simply a result of the sum of the random
fields on all the spins. But when the field is very strong ($\Delta\gg1$),
the rearrangement of field is limited, and, in most cases, the domains
have single spins and the orientation of a spins is in the same direction
as its magnetic field.

The other basic operation is the relabel operation. This operation
updates an auxiliary field, the height, defined for each spin. Following
the more detailed constraints described in Sec.~\ref{sub:Auxiliary-fields-and},
this height guides the pushes. Most simply put, pushes are always
in the downhill direction. When a push is not possible from a given
site, the relabel operation increases the magnitude of the height
of that site. This relabelling will either allow a push to be executed
or will identify the spin as having a particular sign in the ground
state.

The basic operations can be carried out using a variety of ordering
methods. The choice of method affects the running time of the algorithm.
A specific algorithm is defined by two sets of choices, which are
defined and described in detail in the following subsections:

\begin{enumerate}
\item the dynamically-determined order local operations (pushes and relabels),
which is organized by a choice of data structure, and
\item heuristic manipulations of the auxiliary fields, namely, global updates
and gap relabeling.
\end{enumerate}
We implemented the PR algorithm in Java and C++. The Java implementation
can visualize the evolution of the auxiliary fields used by PR. The
C++ code relied on the original C code \cite{GoldbergDownload} developed
by Cherkassky and Goldberg \cite{Cherkassky1997}. The codes can be
downloaded from our web site \cite{RFIMApplet}.

\subsection{\label{sub:Auxiliary-fields-and}Auxiliary fields and push-relabel
operations}

The PR algorithm uses three auxiliary fields to guide the rearrangement
of the magnetic field and to enforce the constraints given by the
bond strengths. One field is the \emph{}residual bond strength (more
commonly referred to as the residual capacity in the literature \cite{Cormen1990}).
This residual interaction between sites is denoted $r_{ij}$. Initially,
$r_{ij}=r_{ji}=J$ for all nearest neighbor pairs $(i,j)$, but in
general $r_{ij}$ need not equal $r_{ji}$ during the execution of
the algorithm. This residual bond strength defines the paths along
which excess magnetic field can be pushed. A site $j$ is said to
be reachable from a site $i$ if there is a directed path from $i$
to $j$ with $r_{kl}>0$ for all bonds $(k,l)$ along that path.

For each site $i$, an excess field $e_{i}$ and a height field $u_{i}$
(which is often called {}``distance'' or {}``rank'') are also
defined. At the beginning of the algorithm, the excess, which can
be positive or negative, is set equal to the random field strength,
$e_{i}=h_{i}$, and the height field $u_{i}$ is set equal to the
distance to the nearest reachable site with negative excess. Sites
from which no site with negative excess can be reached have $u_{i}=\infty$.
Sites with negative excess have $u_{i}=0$. 

The PR algorithm maintains the height field $u_{i}$ in a fashion
designed to move the positive excess {}``downhill'' (to smaller
heights), i.e., towards sites with negative excess, where possible.
If it becomes impossible to rearrange excess towards a site with negative
excess, the site is given a height label $u_{i}=\infty$ (in practice,
$\infty$ is represented by $n$, the number of spins). A site $j$
is said to be accessible from $i$ if the residual bond strength $r_{ij}>0$
and $u_{i}=u_{j}+1$. Any site with positive excess and height $u_{i}<\infty$
is active\emph{.} (In the double queue method mentioned in Sec.~\ref{sub:Data-structures},
negative excess sites can also be active.)

The push operation moves excess from an active site $i$ to an accessible
neighbor $j$. Letting $\delta=\min(r_{ij},e_{i})$, the excesses
are updated by $e_{i}\rightarrow e_{i}-\delta$, $e_{j}\rightarrow e_{j}+\delta$,
while the residual bond strengths are updated according to $r_{ij}\rightarrow r_{ij}-\delta$
and $r_{ji}\rightarrow r_{ji}+\delta$. If an active site has no accessible
neighbor, so that no pushes are possible, it is relabeled: the height
$u_{i}$ is set to one greater than the height of the lowest neighbor
$j$ (i.e., minimal $u_{j}$ over neighbors $j$) for which $r_{ij}>0$.
If no such neighbor exists, the height of $i$ may immediately be
raised to $u_{i}=\infty$. We call the combination of all possible
push operations from a single spin possibly followed by a relabel
a PR step.

The PR algorithm terminates when no active sites remain. The total
number of PR steps needed to complete the algorithm is denoted by
$\NPR$. The assignment of spin orientations in the ground state is
found by executing a global update (Sec.~\ref{sub:Heuristics}) to
finalize which sites have $u_{i}=\infty$. Sites with $u_{i}=\infty$
are assigned $s_{i}=1$, while the remaining spins have $s_{i}=-1$.

\subsection{Data structures\label{sub:Data-structures}}

We implemented the PR algorithm using four different data structures:
a first-in-first-out queue (FIFO), a highest height priority queue
implemented as a heap (HPQ), a lowest height priority queue (LPQ)
also implemented as a heap, and a double FIFO queue (DFIFO) that treats
positive and negative excess symmetrically. (We also tried a stack
or last-in first-out (LIFO) structure, but rejected it due to vastly
longer running times at large $\Delta$.)

The FIFO structure is a list of active sites in the lattice. The site
at the front of the list, $i$, is removed. Excess is pushed away
from $i$ if possible. If any inactive neighbors $j$ of $i$ are
made active through a push operation, they are added to the end of
the list. If there are no accessible neighbors, $i$ is relabeled.
If $i$ is still active after this PR step, it is also appended to
the end of the list.

The HPQ structure \cite{CherkasskyGoldberg97} is more complex than
FIFO. Like the FIFO structure it contains a list of all active sites.
The list is not organized by the temporal order in which sites have
been treated, as would be the case in a FIFO queue, but by the height
of the sites. The first site in the HPQ is always a site with maximal
height. When we remove a site from the front of the queue, the site
with the highest height that is still in the queue moves to the front.
If we add a site with a larger height than any of the sites already
in the queue, the new site moves to the front of the queue. If after
a PR step a site is still active, it is re-added to the queue. Since
a PR step relabels an active site by increasing its height by at least
one before adding it back to the queue, such a site is still of maximal
height and is added at the front. Thus the same site might be acted
upon by the algorithm many times in a row. In practice, the HPQ is
simply implemented by sorting the sites into bins given by their height
values.

The LPQ structure \cite{AVG97} is exactly the same as the HPQ structure,
except that the list is reversed, so that sites $i$ with lowest height
$u_{i}$ are subject to PR operations.

FIFO, HPQ, and LPQ enforce an asymmetry between positive and negative
excess. Why should one sign (positive excess) be pushed around while
the other (negative excess) remains static (except by rearrangement
of positive excess onto a negative excess of lesser magnitude)? Certainly,
the physical problem is unaltered by the replacement $h_{i}\rightarrow-h_{i}$.
Our fourth implementation treats positive and negative excesses on
an equal footing. Negative excess then moves from lower heights (here,
$u_{i}<0$ is a possible height label) to higher heights. We implemented
this as two FIFO queues (DFIFO), one for positive and one for negative
excess sites. Both queues are updated simultaneously. As noted in
the next section, we also used two height fields, one for each sign
of excess, when implementing DFIFO.

\subsection{Heuristics\label{sub:Heuristics}}

If these queues are adapted as described so far, the algorithm, though
polynomial in $n$, is too slow to be practical for studying larger
systems. Heuristics can be used to manipulate the height field to
guide the push-relabel operations. Good heuristics are crucial to
the practicality of the algorithm.

All our implementations use the {}``global update'' heuristic for
initialization of the $u_{i}$. A global update \cite{Cherkassky1997}
is used for two purposes. It creates gradients in the height field
from the sites with positive excess to the nearest site with negative
excess, allowing the excess fields to move efficiently towards annihilation.
It addition, it identifies regions that are disconnected from the
rest of the spins and do not contain any negative excess; such regions
are labeled as spin-up regions and are removed from further consideration.
Without this identification it would take of the order of $nr$ operations
to mark a region of $r$ spins as up. The global update is implemented
as a breadth first search starting from the set of negative excess
sites and takes of the order of $n$ operations on a hypercubic lattice.
During this search, the height of each vertex is set to the distance
to the nearest reachable site with negative excess (sink). One important
result of global updates is that local minima of the height field
with positive height, often residuals of former negative excess sites,
are eliminated at sites with nonnegative excess. Following common
practice, we choose the interval between global updates to be fixed,
with a global update executed after every $\Gamma$ push-relabel (PR)
steps.

The global update needs to be modified for the double queue (DFIFO)
approach. In parallel with the height field for positive excess, based
on a search from nodes with negative excess, it is natural to construct
the height field for negative excess sites using positive excess sites
as {}``sinks''. This creates an ambiguity for sites with zero excess.
Should their height be determined in relation to the positive or negative
excess sites? We did not directly resolve this ambiguity, but instead
used a scheme with separate height fields for the positive and negative
excesses, with separate relabeling and global updates. 

\label{anchor:gap-heuristic}In addition, when using the HPQ data
structure, we use gap relabeling \cite{Cherkassky1997}. If there
is a height $u_{g}$, such that no site has height $u=u_{g}$, but
there are active sites with $u>u_{g}$, i.e., there is a gap in the
set of heights, all sites with $u>u_{g}$ are assigned the maximum
height. This reduces the need for global updates as active sites with
$u>u_{g}$ are disconnected from the sinks. When using HPQ, regions
with larger height tend to have their height raised uniformly, leading
to the possibility of creating a gap quickly. As the gap is a global
tally, it is not efficient at detecting regions that have separated
themselves from their surroundings locally, so gap relabeling is less
useful in, e.g., FIFO, where updates are carried out without respect
to height. To facilitate gap relabeling, the HPQ algorithm uses an
array that contains the number of sites (whether active or not) at
each possible height. A gap is created when the occupation number
at a non-maximal height is reduced to zero.

\section{\label{sec:Timing-Results}Comparison of Implementations}

To find the best available combination of data structure and heuristics
for solving the Gaussian RFIM, we measured the running time of the
algorithm for a variety of combinations. We measure the running time
for the algorithm to find the ground state using both CPU time $t$
in seconds and the total number of PR steps $\NPR$. PR steps are
the core operations of the algorithm and give us a machine independent
measure of the improvement due to the heuristics. However, PR steps
don't account for the time needed for the internal bookkeeping of
the data structure, nor do they reflect the time needed for the heuristics,
such as the global update. The CPU time $t$ is therefore another
useful measure of the performance and ultimately what we want to minimize.
For the timing measurement we used SUN's Java Virtual Machine 1.4.2
on Dual 1GHz PIII machines with 512MB of RAM running Linux.

Here we summarize the results of our timing runs. We first describe
our results for two-dimensional lattices. In this case there is no
phase transition for the RFIM ground states at finite $\Delta$. However,
as the correlation length $\xi$ diverges very rapidly as $\Delta$
decreases, there is a crossover disorder value $\Delta_{x}(L)$ at
which $\xi$ exceeds the linear system size $L$. For $\Delta<\Delta_{x}(L)$,
the 2D system is effectively ferromagnetic. The dependence of $\Delta_{x}$
on $L$ is very slow \cite{Binder1983,GrinsteinMa82,SeppalaAlava2D2001}.
In our simulations, we find $\Delta_{x}(8)\approx1.7$ and $\Delta_{x}(4096)\approx0.55$.
In three dimensions, there is a true transition at $\Delta=\Delta_{c}\approx2.27$
\cite{MiddletonFisherRFIM}.

In this section, we compare the timings for FIFO and HPQ structures
and then present results for DFIFO. (Results for LPQ are included
in the discussion of the results for $d=3$, near the end of this
section.) When seeking to minimize the running time, there is a competition
between frequent global updates (small $\Gamma$), which improve the
efficiency of the PR steps and infrequent updates (large $\Gamma$),
which save the cost of performing a global update. We first determine
the global update interval that minimizes the CPU time over all choices
of $\Delta.$ Generally, for all data structures, we find a minimum
in $t$ for $\Gamma_{\min}\approx n$ (or at least that performance
is not improved for other $\Gamma$). The FIFO and HPQ structures
then give comparable results away from the crossover $\Delta\approx\Delta_{x}$,
though FIFO is preferable at small $\Delta$ and HPQ is faster at
large $\Delta$. DFIFO takes significantly more time than either HPQ
or FIFO to find the ground state for $\Delta$ on the order of $\Delta_{x}$.
We also find that the discreteness of global updates can lead to plateaus
in the number of global updates $N_{G}$ as a function of $\Gamma$.
The 3D results are generally consistent with our results for 2D lattices.
We find that LPQ always performs at least as well as HPQ and is faster
for $\Delta>\Delta_{c}$.

\subsection{Choosing the global update interval in two dimensions}

We first examine the effect of varying the update interval $\Gamma$
for two dimensional samples. Global updates are only useful if they
reduce the number of PR steps needed to find the solution. As we mentioned
in Sec.~\ref{sub:Heuristics}, global updates are expensive, taking
of the order of $n$ operations, and should therefore not be performed
too frequently. To find the optimal update interval $\Gamma_{\min}$,
we varied $\Gamma$ for each data structure over several orders of
magnitude for various values of the random field strength $\Delta$
and examined how both $\NPR$, the number of push-relabel cycles used
to find the ground state, and the running time $t$ varied.

\subsubsection{Global update interval for FIFO }

Immediately after a global update, push operations are guaranteed
to move excess towards the nearest sinks. After a global update, however,
some sinks are soon annihilated by excess pushed into them. If a sink
at site $s$ is annihilated, the height field around $s$ no longer
indicates the shortest distance to a sink; the height field still
slopes towards $s$ although there is no longer a sink. The extent
of this {}``misleading'' height field depends on the density of
sinks and is limited by the distance between sinks. When $\Gamma$
is small, global updates are frequent enough that the height field
generally leads positive excess towards a sink. As long as this is
the case, increasing the frequency of global updates does little to
reduce $\NPR$. Therefore, for small $\Gamma$, we expect $\NPR$
to be almost independent of $\Gamma$ for FIFO.

\begin{figure}
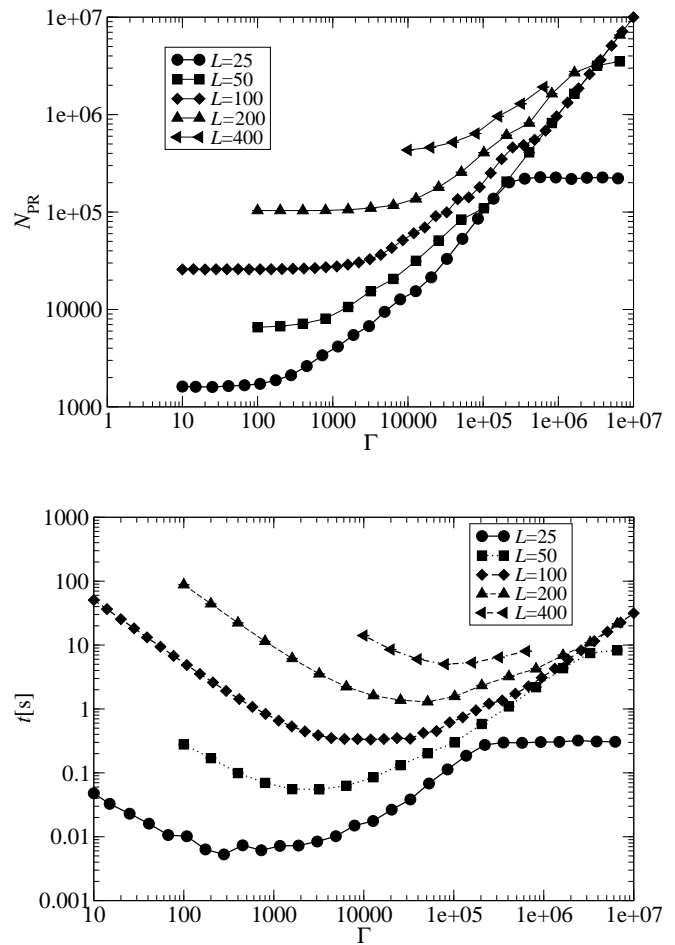

\begin{center}\includegraphics[%
  width=1.0\columnwidth,
  height=0.35\textheight,
  keepaspectratio]{15bQ.eps}\end{center}

\begin{center}\includegraphics[%
  width=1.0\columnwidth,
  height=0.35\textheight,
  keepaspectratio]{15aQ.eps}\end{center}

\caption{\label{cap:Times-for-Queue}Mean running time for finding the ground
state of a 2D RFIM system vs.~$\Gamma$ using FIFO and $\Delta=2.2$,
where the correlation length is much smaller than the sample sizes
used, $\xi\ll L$. For clarity in this figure and other figures in
this section, statistical error bars are not shown, but are consistent
with the apparent deviations from smooth curves. The figure shows
(a) the number of PR steps $\NPR$ used to find the ground state and
(b) CPU time $t[\textrm{s}]$ as a function of the global update interval
$\Gamma$ for $L=25$, 50, 100, 200, and 400. For small $\Gamma$,
$\NPR$ is nearly independent of $\Gamma$ showing that frequent global
updates are unnecessary. At $\Gamma$ somewhat less than $0.1n=0.1L^{2}$,
$\NPR$ starts to increase. The minimum in the CPU time $t$ is at
$\Gamma_{\min}\approx n$, where the change in time needed for the
additional PR steps balances the change in time needed for the global
update. }
\end{figure}

\begin{figure}
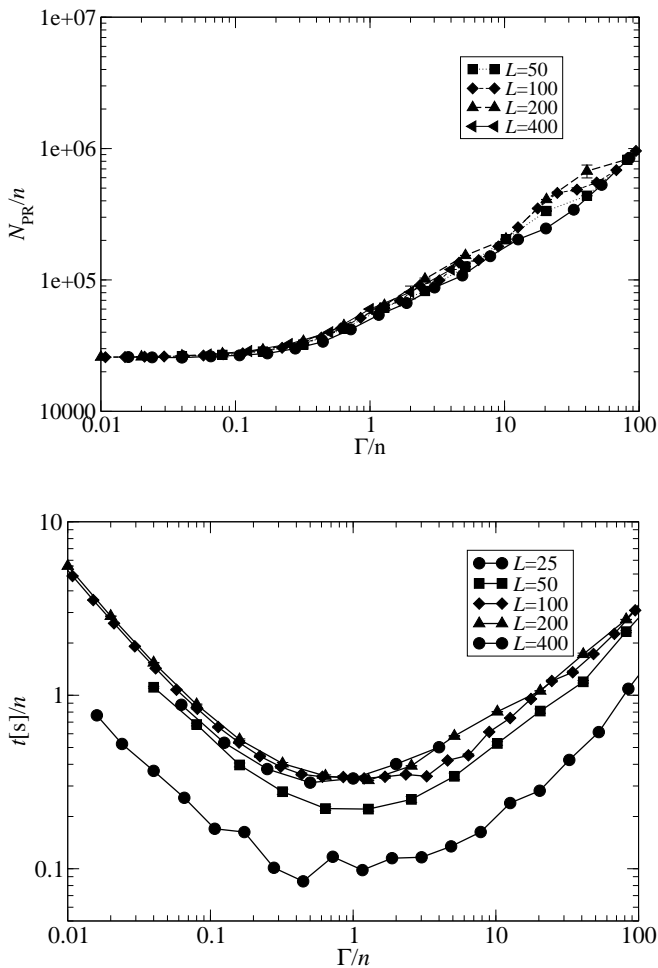

\begin{center}\includegraphics[%
  width=1.0\columnwidth,
  height=0.35\textheight,
  keepaspectratio]{15bQscaled.eps}\end{center}

\begin{center}\includegraphics[%
  width=1.0\columnwidth,
  height=0.35\textheight,
  keepaspectratio]{15aQscaled.eps}\end{center}

\caption{\label{cap:Scaling_for_queue}Scaling of PR steps $\NPR$ and CPU
time $t$ with system size $n$ for FIFO ($\Delta=2.2$). (a) The
average number of PR steps per site, $\NPR/n$, as a function of number
of push-relabel steps per site per global update, $\Gamma/n$. The
data collapse is consistent with the number of operations scaling
with $n$ (the best that can be expected for $\xi\ll L$). (b) CPU
time $t$ per site ($t/n$) vs.~$\Gamma/n$. The values along both
axes have been divided by $n$ the number of sites. \textcolor{magenta}{}A
fair collapse develops for $L\ge100$. The minimum is at $\Gamma_{{\rm min}}/n\approx1$,
independent of system size. The minimum is shallow, however, and even
missing $\Gamma_{{\rm min}}$ by a factor of ten increases the running
time only by a factor of about two.}
\end{figure}

Fig.~\ref{cap:Times-for-Queue} displays algorithm costs $\NPR$
and $t$ for the case $\Delta=2.2$, which exceeds $\Delta_{x}$.
The expectation that $\NPR$ is independent of $\Gamma$ at small
$\Gamma$ is confirmed by the plot of the mean of $\NPR$ as a function
of $\Gamma$ displayed in Fig.~\ref{cap:Times-for-Queue}(a). As
we increase $\Gamma$, a larger and larger fraction of the height
field becomes incorrect between global updates: more minima of the
height field no longer contain sinks. The plot shows that $\NPR$
starts to increase significantly above $\Gamma\approx0.1n$. To minimize
the running time $t$, we need to balance the cost of additional global
updates with the reduction in $\NPR$. A global update takes of the
order of $n$ operations and, in fact, Fig.~\ref{cap:Times-for-Queue}(b)
shows a minimum in $t$ vs.~$\Gamma$ at $\Gamma_{{\rm min}}\approx n$.
To verify our assumption that $\Gamma_{{\rm min}}\approx n$, we rescaled
the curves in both Fig.~\ref{cap:Times-for-Queue}(a) and (b) by
dividing $\Gamma$, $\NPR$, and $t$ by $n$. The collapse of the
data verifies the minimum in $t$ vs.~$\Gamma$ at $\Gamma_{{\rm min}}\approx n$
and shows that the running time of the algorithm scales nearly linearly
with $n$ in 2D (Fig.~\ref{cap:Scaling_for_queue}), at fixed $\Delta$.

\subsubsection{\label{sub:detailed_look}Detailed look at the FIFO data}

While determining the optimal global update interval for the FIFO
structure, we noticed two features of interest in the data. These
features are the tendency for $\NPR$ to be an integer multiple of
$\Gamma$ and, at small $\Delta$, a separation of the mean run times
between positively and negatively magnetized samples.

\begin{figure}
\begin{center}\includegraphics[%
  width=1.0\columnwidth,
  keepaspectratio]{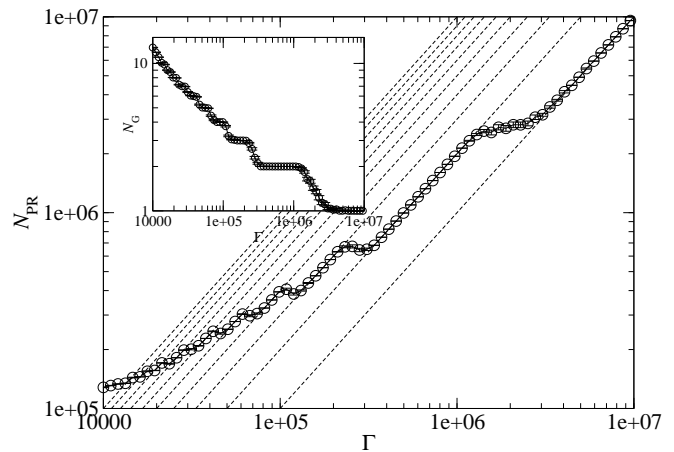}\end{center}

\caption{\label{cap:-y=3Dnx-}Plot showing the piecewise linear behavior of
$\NPR$, the mean number of PR steps needed to find the ground state,
when plotted as a function of $\Gamma$, the interval between global
updates, for FIFO in $d=2$. The parameter values are $L=200$ and
$\Delta=2.2$ and the data is averaged over 100 samples. $\NPR$ increases
linearly over intervals in $\Gamma$, $\NPR=k\Gamma$, with the curves
for $k=1,2,3,\ldots$ indicated by the dashed lines. The large changes
caused by the global update operation can lead to plateaus in the
number of global updates $N_{G}$ vs.~$\Gamma$, as shown in the
inset.}
\end{figure}

A close look at the data reveals piecewise linear behavior in the
average running time of the algorithm, when measured by $\NPR$, as
displayed in Fig.~\ref{cap:-y=3Dnx-}. These linear regions are consistent
with $\NPR=k\Gamma$, for integer $k$, as shown by the dashed lines
in the figure. These linear regions coincide with plateaus in the
number of global updates executed during the solution, $N_{G}$, when
plotted vs.~$\Gamma$ (see the inset in Fig.~\ref{cap:-y=3Dnx-}).
The plateaus in $N_{G}$ vs. $\Gamma$ reflect the effect of the global
update, which causes large changes in the height field and can bring
the auxiliary fields close to a solution. Note that a ground state
is found by the algorithm when all of the positive excess is confined
to regions that are isolated from sinks. This isolation is due to
saturated bonds that block the rearrangement of flow (and to the cancellation
of positive excess with negative excess in regions accessible to the
sinks). The algorithm will not terminate until the blocked-off regions
have their heights raised to $u_{i}=n$. Without global updates ($\Gamma=\infty$),
there is a separation between the times when the push-relabel operations
effectively determine the domain boundaries by saturating the appropriate
bonds and the time when the relabel operations identify the up-spin
regions. When $\Gamma$ is finite, but larger than the number of PR
steps needed to find all of the domain boundaries and smaller than
the total number of PR steps needed to terminate the algorithm, the
first global update effectively terminates the running of the algorithm
and $N_{G}=1$. There is another interval at smaller values of $\Gamma$
where one global update is executed before the domain boundaries are
determined and the second global update terminates the algorithm,
giving $N_{G}=2$. This pattern continues to higher values of $m$.
We emphasize that the data presented is averaged over $100$ samples
at each value of $\Gamma$; the data therefore indicates that the
fluctuations in the locations of the linear regions are quite small.

We also noted a strong up-down asymmetry in the running time at small
$\Delta$. For $\Delta\ll\Delta_{x}$ and $\Gamma=n$, it takes about
50\% longer to find the solution if the ground state is spin-up than
if it is spin-down. For $\Gamma\rightarrow\infty$ the ratio of mean
running times becomes very large. At small $\Delta$, the ground state
is determined by the sum over all random fields. If the sum is negative,
the algorithm is done as soon as all the positive excess fields have
been annihilated, but if the sum is positive all the sites with remaining
excess fields must be moved up to height $n$ before they are labeled
inaccessible. This local relabeling can take of the order of $n^{2}$
steps, as essentially each site must be moved stepwise to the maximal
height. When a global update is included, the height label of all
sites is set to its maximum value once no more sites with negative
excess exist. As the global relabeling takes only of the order of
$n$ steps, this makes the running times for up- and down- magnetized
samples more similar, to within $O(n)$ in total magnitude, for $\Gamma=O(n)$.

\subsubsection{Global update interval for HPQ}

\begin{figure}
\begin{center}\includegraphics[%
  clip,
  width=1.0\columnwidth]{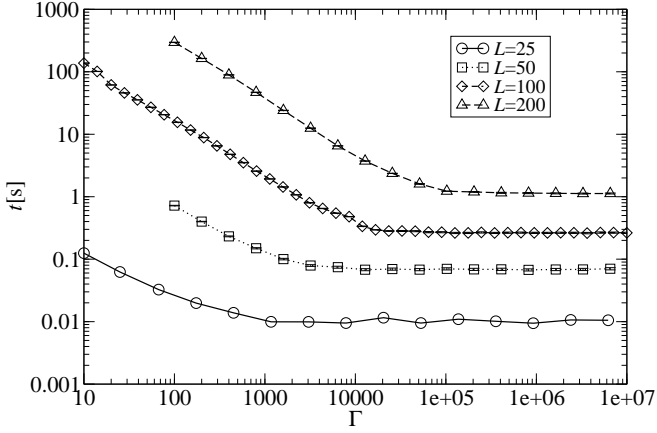}\end{center}

\caption{\label{cap:Time-for-Heap}CPU time $t[\textrm{s}]$ for HPQ with
gap relabeling vs.~$\Gamma$ for $\Delta=2.2>\Delta_{x}$. In this
regime gap relabeling is effective and global updates are almost unnecessary.
For $\Gamma<n$, the running time increases due to frequent global
updates. For $\Gamma>n$, the running time stays constant. There is
no discernible minimum in the $t$ vs.~$\Gamma$ curve, but $\Gamma=n$
does not hurt performance at large $\Delta$ and improves performance
for smaller $\Delta$.}
\end{figure}

For $\Delta<\Delta_{x}$, the behavior of the timing for HPQ vs.~$\Gamma$
is very similar to the behavior of the timing for FIFO, with $\Gamma_{{\rm opt}}\approx n$.
In contrast, in the paramagnetic regime, gap relabeling detects enclosed
domains efficiently and quickly reduces the number of active sites.
For large $\Delta$, then, frequent global updates ($\Gamma<n)$ quickly
dominate the running time $t$. For $\Gamma>n$, $t$ becomes independent
of $\Gamma$ (Fig.~\ref{cap:Time-for-Heap}), as global updates are
not executed before the solution is found.%
Consistent with the optimal behavior for small $\Delta$ and the independence
of $t$ from $\Gamma$ at large $\Delta$, we will use $\Gamma_{{\rm opt}}\approx n$
for the HPQ structure. As $\Gamma=\infty$ is a reasonable choice
for larger $\Delta$, we will sometimes include this choice for comparison.

\subsection{FIFO vs. HPQ in two dimensions}

Now that we have found an optimal value for $\Gamma$ for each data
structure, we can directly compare the performance of FIFO and HPQ
for various $\Delta$. We compare timings using $\Gamma=n$ for FIFO
and both $\Gamma=n$ and $\Gamma=\infty$ for HPQ. We will refer to
the latter two choices as HPQ$_{n}$ and HPQ$_{\infty}$, respectively.
We computed the mean values of $t$ and $\NPR$ for a large range
of system sizes and disorders. The running times were averaged over
at least 100 (sometimes as many as $10^{5}$) samples, chosen independently
for each value of $L$ and $\Gamma$.

\begin{figure}
\begin{center}\includegraphics[%
  width=1.0\columnwidth]{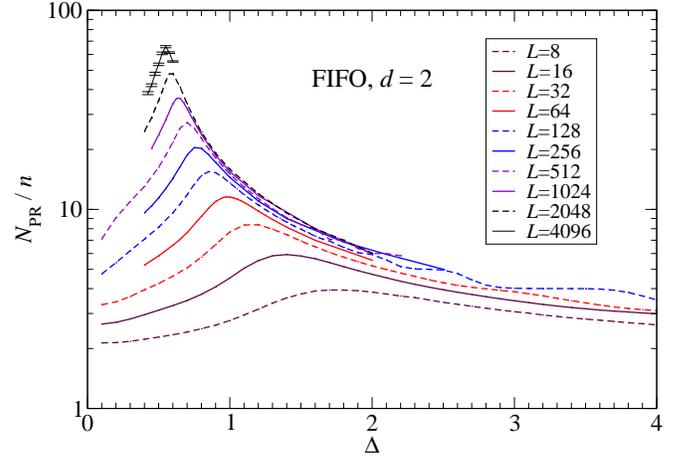}\end{center}

\caption{\label{cap:Critical-slowing-down-in-2d}{[}Color online{]} Number
of push-relabel operations per site needed to solve for the ground
state, $\NPR/n$, plotted as a function of $\Delta$, for a 2D square
lattice with Gaussian disorder. The number of spins $n=L^{2}$ ranges
from $8^{2}$ to $4096^{2}$. Due to the large range of system sizes
and hence running times, we have used a logarithmic axis for plotting
$\NPR/n$. Statistical error bars and individual points are not displayed,
as the point spacing is small and the statistical uncertainties are
very small, except for $L=4096$. The global update interval used
was $\Gamma=n$. Plateaus are seen when $\NPR$ is close to a multiple
of $\Gamma$, especially at larger $\Delta$. Each curve show a pronounced
peak, which can be used to define $\Delta_{x}(L)$. At this value
of $\Delta,$ the correlation length is of the order of the system
size and the system crosses over from a ferromagnetic regime with
one large domain to a paramagnetic regime where the correlation length
is smaller than the system size. The crossover field $\Delta_{x}$
approaches zero as L goes to infinity, but the approach is logarithmically
slow.}
\end{figure}

\begin{figure}
\begin{center}\includegraphics[%
  width=1.0\columnwidth]{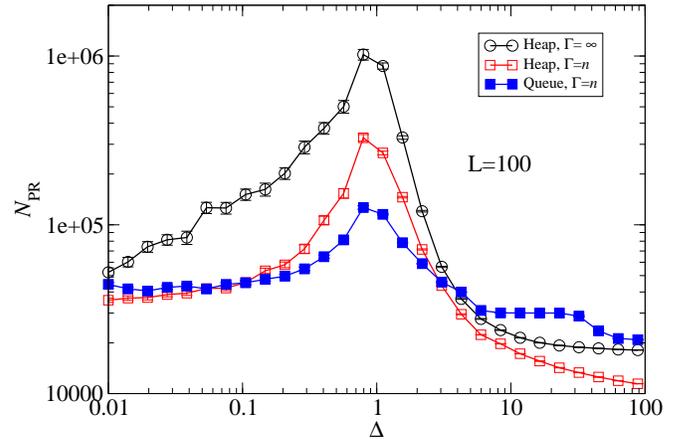}\end{center}

\caption{\label{cap:TimingGeneral2D}{[}Color online{]} Number of PR steps
$\NPR$ in $d=2$ for $L=100$. This plot shows the timing for HPQ
with ($\Gamma=n$) and without ($\Gamma=\infty$) global updates,
compared with FIFO. Note the significant improvement the global update
gives to HPQ for field strengths below and at the finite-size crossover.FIFO
needs fewer push operations than HPQ$_{\Gamma=\infty}$ over almost
3 orders of magnitude in $\Delta$. This difference becomes more pronounced
as the system size increases.}
\end{figure}

Fig.~\ref{cap:Critical-slowing-down-in-2d} displays the running
time per spin for FIFO with $\Gamma=n$ and Fig.~\ref{cap:TimingGeneral2D}
presents a comparison of all three combinations for $L=100$. %
The first notable feature in the data is that all three combinations
show a pronounced peak in $\NPR$ at a value of $\Delta_{x}(L)$ between
0.5 and 2.0 over the range $L=8$ to $4096$. This peak is reminiscent
of the critical slowing down near a phase transition. It moves slowly
to the left with increasing system size (Fig.~\ref{cap:Critical-slowing-down-in-2d}),
consistent with $\Delta_{x}(L)$ decaying with increasing $L$. As
noted in Sec.~\ref{sec:Random-field-Ising}, there is no phase transition
in 2D, but there is a crossover for any finite size system from a
ferromagnetic regime at low $\Delta$ to a paramagnetic regime at
large $\Delta$ as the correlation length becomes smaller than the
system size. 

For small $\Delta$, HPQ$_{n}$ and FIFO perform very similarly. HPQ$_{\infty}$,
on the other hand, is 2 to 4 times slower in this regime. Near the
crossover, the running times for HPQ$_{n}$ and FIFO start to deviate
and the difference is largest at $\Delta_{x}$, where HPQ$_{n}$ is
about 3 times slower than FIFO and HPQ$_{\infty}$ is yet another
factor of 3 slower than HPQ$_{n}$.

For $\Delta>\Delta_{x}$, FIFO starts to lose its advantage as domain
sizes become smaller and gap relabeling becomes more and more effective
with increasing $\Delta$. When $\Delta\approx3$, HPQ$_{n}$ starts
to outperform FIFO; when $\Delta\approx20$, FIFO takes twice as many
push operations as HPQ$_{n}$.

\subsection{Timings for the double Queue (DFIFO)}

Our method for simultaneously rearranging both positive and negative
excess, DFIFO, performs well for small $\Delta$ in $d=2$ but never
better than FIFO. For large values of $\Delta$, it is much slower
than FIFO and HPQ. DFIFO performs even worse in 3D. The reason became
clear when we visualized the rearrangements of excess field. The visualization
shows that positive and negative excesses tend to miss each other.
The potential landscape, given by the two height fields, is not updated
when excess is pushed and is not consistently updated by relabels.
The rearrangement of excess is generally guided by the last global
update: the negative excess is pushed to where the positive excess
was at the time of the last global update, and vice versa. This lack
of coordination between the two height fields becomes more pronounced
in higher dimensions, where there are more possible paths between
sites. It is possible that using some other combination of the height
fields for a heuristic would produce better results. \textcolor{magenta}{}One
alternative, for example, would be to combine the separate height
fields for positive and negative excesses into a single field. Positive
excess would move down the height field gradient and negative excesses
would move up the same gradient.

\subsection{\label{sub:Results-for-three}Results for three dimensions ($d=3$)}

In three dimensions, $\Gamma=n$ remains a good choice for both FIFO
and HPQ. As in the case $d=2$, due to a very broad minimum in the
dependence of $t$ on $\Gamma$, the exact value of $\Gamma$ is not
critical. This is in agreement with previous results where $\Gamma_{\mathrm{min}}=n$
has been found to be a good choice for a variety of sparse and dense
graphs \cite{Cherkassky1997}. 

\begin{figure}
\begin{center}\includegraphics[%
  width=1.0\columnwidth,
  keepaspectratio]{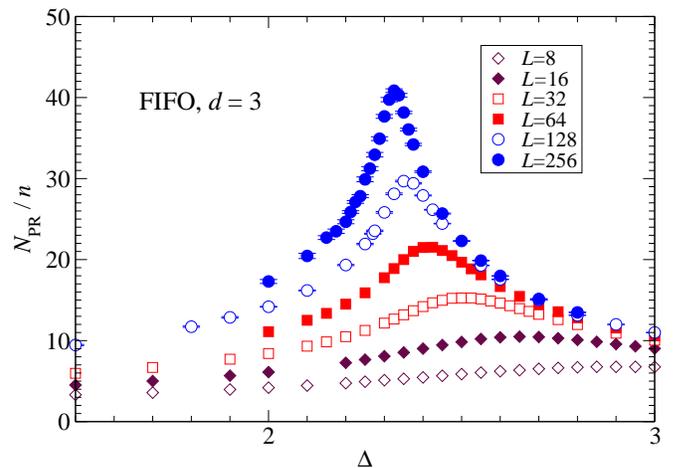}\end{center}

\caption{\label{cap:Scaled-Number-of-Pushes-3d}{[}Color online{]} Push-relabel
operations (cycles) per site, $\NPR/n$, plotted vs.~disorder strength
$\Delta$ in $d=3$, for $L=8$, 16, 32, 64, 128, and 256, for the
push-relabel algorithm using the FIFO data structure. The global update
interval is fixed at $\Gamma=n$. There is a clear peak in the running
time near $\Delta\approx2.3$, which is in good agreement with the
critical field $\Delta_{c}=2.270(5)$ found \cite{MiddletonFisherRFIM}
for the ferro- to paramagnetic phase transition for a Gaussian distribution. }
\end{figure}

\begin{figure}
\begin{center}\includegraphics[%
  width=1.0\columnwidth,
  keepaspectratio]{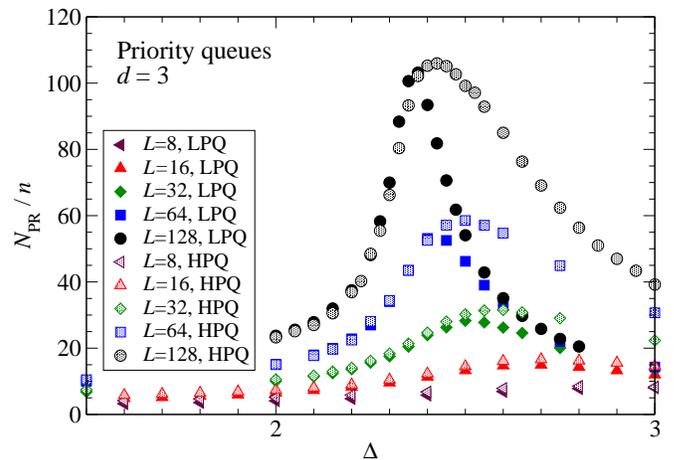}\end{center}

\caption{\label{cap:HPQ-LPQ-3D-raw}{[}Color online{]} Push-relabel operations
(cycles) per site $\NPR/n$ vs.~$\Delta$ in $d=3$, for $L=8$,
16, 32, 64, 128, and 256, when using the HPQ and LPQ data structures.
There is again a clear peak in the running time near $\Delta\approx2.3$,
but the peak is significantly broader and higher for HPQ, than for
LPQ. Note that the number of cycles grows more quickly with $L$ than
for the FIFO structure.}
\end{figure}

The peak in the timing of the algorithm is quite pronounced, as in
the 2D data, but in this case the location of the peak converges to
a fixed value at large $L$. We gathered data for the number of push-relabel
operations $\NPR$ per site, $\NPR/n$, vs.~$\Delta$, for sizes
$L$ ranging from 8 to 128 for HPQ and LPQ and up to size 256 for
the FIFO data structure. The plot of the FIFO data (Fig.~\ref{cap:Scaled-Number-of-Pushes-3d})
shows a growth in $\NPR/n$ with $L$ and convergence to a well-defined
curve for the running time for $\Delta$ in the paramagnetic range.
In the ferromagnetic regime, there is a slow growth of the running
time with sample dimension $L$. Both the LPQ and HPQ data structures
are significantly slower than the FIFO data structure, at moderate
values of $\Delta$ (with a ratio of $\approx3$ for the peak running
times at $L=128$), as seen in Fig.~\ref{cap:HPQ-LPQ-3D-raw}. The
LPQ data structure is significantly faster than the HPQ structure
on the paramagnetic side of the peak in the running time, but is very
similar in speed on the ferromagnetic side. The running time for LPQ
converges much more quickly than the HPQ version to an $L$-dependent
value as $L$ is increased at fixed $\Delta>\Delta_{c}$.%

\section{Qualitative Description\label{sec:Qualitative-Description}}

\begin{figure}
\begin{center}\includegraphics[%
  width=1.0\columnwidth,
  keepaspectratio]{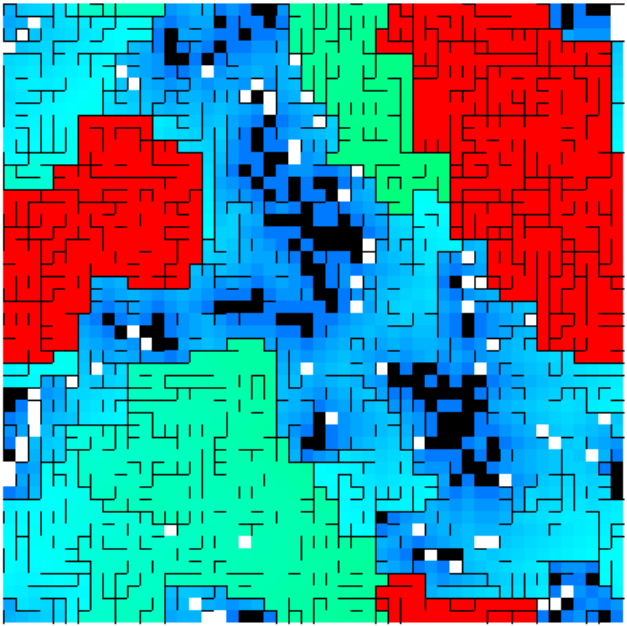}\end{center}

\caption{\label{cap:Partial-solution}{[}Color online{]} Image of the state
of the auxiliary fields during the solution of a 2D RFIM of size $L=50$
and with disorder strength $\Delta=2$. This picture shows a snapshot
of the height field $u_{i}$, the location of non-zero excess $e(i)$,
and bonds with zero residual strength ($r_{ij}=0$), while executing
the push-relabel algorithm (HPQ data structure). The red (darkest
non-black) regions have already been identified as up; the heights
in this region have the maximal value $n=2500$. The colors that are
not black or white correspond to the height field in the remainder
of the sample: blue (darker colors) indicates a low, while green (lighter
colors) indicates a high height. The white squares represent spins
with a positive excess field (the height at these sites is not indicated);
black squares represent spins with a negative excess (and have height
zero). The black lines are dual to the bonds that are saturated (i.e.,
have zero residual strength). These saturated dual bonds ($r_{ij}=0$
or $r_{ji}=0$) {}``block'' the rearrangement of excess $e_{i}$.
More up-spin regions may be identified as more bonds are saturated
and the dual bonds link up to isolate a region. The algorithm terminates
when all positive excess is confined to up-spin regions.}
\end{figure}

In order to better understand the timing results, we have visualized
the evolution of the height fields and the rearrangement of excess.
The visualization code \cite{RFIMApplet} uses a color map to display
the height field. The program has the option to display the location
of sites with excess (white for positive and black for negative) and
to indicate where the flow is saturated, i.e., where $r_{ij}=0$.
A sample snapshot from the simulation with both of these options activated
is shown in Fig.~\ref{cap:Partial-solution}. %

When $\Delta$ is somewhat larger than $\Delta_{x}$ in $d=2$, the
differences between the temporal progress using the HPQ and FIFO data
structures are clearly seen in the dynamic visualization. As FIFO
cycles through all active sites, all the positive excesses move at
a roughly uniform speed down the height gradients. The utility of
the global update at late times for FIFO is apparent: when $\Gamma$
is large (infrequent global updates), a few positive excesses are
seen to skate around in regions contained by saturated bonds. The
pushes have found the minimum cut by saturating the bonds that separate
the positive and negative spin regions, but the algorithm hasn't confirmed
that fact yet. The local relabels, which tend to raise a site's height
only by one step at a time, are inefficient in raising a large up-spin
domain to maximal height. The remaining positive excesses are shuffled
around within the domain, slowly increasing the height by small relabels.
When global updates are infrequent, the algorithm may terminate by
a final global update, which finds that a set of positive excesses
is isolated from all sinks. This confirms the picture discussed in
Sec~\ref{sub:detailed_look}. HPQ, on the other hand, tends to act
repeatedly on the same site. Isolated regions with positive excess
are raised uniformly above the rest of the sample. This allows the
gap heuristic to quickly identify isolated positive spin domains,
even when $\Gamma$ is large.

For samples with weak disorder, the algorithm with the HPQ data structure
does lead to a few positive excesses racing around the network while
the others sit idly. The time to raise a large positive domain high
enough to create a gap is large. Again, active sites tend to move
around quite a bit, while other regions remain unchanged. Towards
the middle of the algorithm execution, the lattice often displays
lines or rings of positive excess sitting on an equipotential line
near a sink, often immediately adjacent to it. These structures are
at low height and cannot move until all excesses of greater height
have been removed. This should slow the algorithm down. The positive
excesses do not {}``screen'' the sinks, since the global update
doesn't differentiate between sinks of different capacity. This expectation
is consistent with our results for the LPQ, which performs somewhat
better than HPQ, especially at higher field strengths.

At large field strengths, the proportion of active sites whose random
field strength is greater than the strength of their bonds to neighbors
(i.e., $\Delta>4$ in 2D) is large. The algorithm acts on most sites
only twice. One can clearly see how HPQ sweeps the lattice and raises
(nearly) all the sites of initial height two (those with no adjacent
sinks) to the maximum height, then does the same to sites of height
one. FIFO also generally makes two or three passes, though its first
pass executes pushes on the sites where pushing is possible and relabels
otherwise.

\section{Summary\label{sec:Summary}}

As has been demonstrated before \cite{Bastea1998a,Duxbury2001,Esser1997,Hartmann1998,Hartmann1998a,MiddletonFisherRFIM,Ogielski1986,Seppala1996},
the PR algorithm is an efficient method to find the exact ground state
for the RFIM at T=0 in any dimension. Here we investigated how to
implement PR to find the ground state most quickly. We compared a
number of data structures and the effect of global and gap relabeling
on the performance of the algorithm. 

In agreement with previous work \cite{Seppala1996}, our detailed
results recommend the FIFO-queue combined with global updates every
$\Gamma\approx kn$ steps, with $k$ a number near unity. FIFO performs
much better near the crossover disorder ($\Delta_{x}$ for $d=1,2$)
and the critical disorder ($\Delta_{c}$ in $d=3$) and never performs
significantly worse than HPQ. The exact value of $\Gamma$ is not
crucial since the minimum in $t$ vs.~$\Gamma$ is very broad.

We also tried an implementation that treats positive and negative
excesses equally. This implementation, however, suffers from the lack
of coordination between the height fields for the two sets of excesses
and positive and negative excesses tend to miss each other. It is
likely that this algorithm could be improved by using a single height
field to coordinate the motion of the positive and negative excess.

A more flexible approach to the global update interval might also
be useful in speeding up simulations. It is likely that adaptively
modifying the global updates so that they are executed when the sink
density changes by a defined fraction or packets of excess have travelled
a given distance would optimize the algorithm during each stage of
the solution. There is still a lot of room for other modification,
e.g., cutting off the breadth first search to reflect saturation and
non-uniform intervals for global update that depend on the number
of sinks that have been annihilated since the last global update rather
than the number of PR steps.

We found that the visualization of the operations (see \cite{RFIMApplet}
for the source code) greatly improved our understanding of the algorithm.
This code may be useful in suggesting further improvements to the
algorithm.

\begin{acknowledgments}
This work has been supported by the National Science Foundation under
grants ITR DMR-0219292 and DMR-0109164.

\bibliographystyle{apsrev}
\bibliography{./HMM}
\end{acknowledgments}

\end{document}